# Correct usage of transmission coefficient for timing the market

Ovidiu Racorean

e-mail: decontatorul@hotmail.com

**Abstract**

Traders and investors involved in an option contract having the underlying stock in range bound are likely to lose their initial investment. Timing in buying an option contract is of capital importance. In a recent article [1] the hypothesis of range bound market is used in conjunction to Black-Scholes equation to find the transmission coefficient relation that help market professionals to correctly timing their investment and risk taking decisions. The present paper explores the theoretical basis of transmission coefficient and its empirical evidence on the market.

1. **Introduction**

Market practitioners are not particularly fans of range bound markets. Prices are moving between the support and resistance levels not trending in any direction. Still, as the statistics revels the markets are in range bound almost 85 percent of the time, dominating the life of traders.

For the option practitioners is more a bad news, taking into account the strike price that has to be surpassed and the life time of a contract. The market professionals involved in an option contract having the underlying stock in range bound is almost certain that their initial investment is history. It is obvious that the timing of buying an option contract is of capital importance.

Another aspect involving the importance of timing comes from the profit side of American options. Having the possibility of being exercised at any time, up to the maturity, traders can make significant profits if correctly timing the short big moves in the stock market.

Exactly this kind of range bound market problems are addressed in the recent paper "Time independent pricing of options in range bound markets" [1]. The paper deals with Black-Scholes equation emphasizing a range bound behavior for the underlying option stock price. Under this hypothesis a time independent equation for pricing the options is derived.

The reasoning is taken further and the probability of price penetrating out of the bounded region, called transmission coefficient is deduced. The transmission coefficient is the effect of a particular price movement in the market characterized by an explosive penetration of support or resistance level named- tunneling effect.

Empirical evidences of such tunneling effect, which in parenthesis being told, is very often encountered in trading practice, are shown in a further section of the paper.

Examples of computing transmission coefficient, distance of price penetration and the short time dramatic fall in volatility for LNKD, GOOG, HUM, NFLX stocks are discussed along with the steps market professionals must take to correctly implement the trading strategy.

2. **Theoretical basis for transmission coefficient**

As it could be expected, talking about option contracts, the Black-Scholes valuation formula must come into discussion. The Black-Scholes equation stays at the foundation of price tunneling effect. The paper [1] that stays at the foundation of the present attempt to construct an option winning strategy explains in details the mathematical deduction process of the relations that would come in discussion in the following sections. The reader is encouraged to check the mentioned paper to better understand the concepts that will be used in constructing the trading strategy.

Not getting deeper in technical description of the tunneling effect, the principal ideas must be briefly exposed to better understand not just the mathematical aspects but also its market applicability.

The Black-Scholes equation suffers a separation of variables in the hypothesis of range bound markets and a time independent equation is deduced in the form:

$$-\frac{\sigma^4}{r(\sigma^2+r)}\frac{d^2\psi_{(S)}}{dS^2} + \frac{1}{S^2}\psi_{(S)} = \frac{r}{\sigma}\psi_{(S)} \qquad (1)$$

The equation simply defines the price movements between levels of support and resistance and in some particular cases, the explosive bounces of price out of this bounded region. These particular cases are related with values of parameters risk-free interested rate r and volatility σ that appear in the right hand side of the equation (1). A brief discussion of the values of the two parameters is necessary here. For simplicity the terms in the right hand side of the equation is noted with **λ**.

$$\lambda = \frac{r}{\sigma} \qquad (2)$$

The values of **λ** are determinant further in discussing the probability of price penetration out of the range bound. For now just have to be stressed out that the bigger the **λ** value is, the most probable an explosive tunneling out of the bounded region is, as the equation (1) is require.

In order to be at high levels **λ** must be situated in one of below cases:

1. Risk-free interest rate is high;
2. Volatility is at low levels.

The interest rate is relatively stable along the life of an option contract so that the stock volatility will remain in the principal role. To recap I should say that in order to see an explosive jump of stock price out of supply or resistance levels the volatility have to fall at low levels in short period of time. Under these circumstances the price of a sock will effectively tunnel out of range bound – exploding higher or lower- and make the option value become for a certain period of time much underappreciated.

Traders can take advantage of these brutal price movements as it will be seen in a further section. To take advantage of the price tunnel effect market professionals should know the probable time of the event will happen. The probability of price penetrating out of the range bound – called transmission coefficient-has long and complicated mathematical ongoing, involving elements of quantum mechanics. The interested reader is encouraged to check it on [1]. Here is enough to mention only the relation of the transmission coefficient:

$$T = e^{-2\sqrt{\frac{r}{\sigma^4}(\sigma^2+r)}\left[\frac{1}{2}\ln\left(\left|\frac{\sqrt{1-\frac{r}{\sigma}K^2}+1}{\sqrt{1-\frac{r}{\sigma}K^2}-1}\right|\right) - \sqrt{1-\frac{r}{\sigma}K^2}\right]} \qquad (3)$$

where K is the distance between resistance and support levels. The relation may look complicated at first sight but is extremely easy to be computed since all its elements are well known by traders.

As market practice shows a transmission coefficient T higher than 95% make a price tunneling effect highly probable.

### 3. Empirical evidences of price tunneling effect

I already stated in latter sections that the tunneling price move is not at all a rare event in the market and once again I confirm that here. For the thousands of stocks that are traded every day some of them will experience tunneling effect in a daily basis. I am convinced that there is no trader who has not saw many times brutal moves with high magnitude of stock price out of the range bound. That makes me not to insist on this aspect of stock market and just show some recent examples of price tunneling effect in Figure 1 below:

Figure 1

I insisted to find charts having attached not only the stock price evolution but also the volatility to evidence along with the explosive move of price out of the range bound, the dramatic fall in volatility I stated in the last section.

I think Figure 1 captures very well both the desired phenomena, the price tunneling effect and the fast fall of volatility to low levels.

### 4. Trading by transmission coefficient

In timing the best moment to be involved in trading the market, transmission coefficient is a great tool. A market could be in range bound for an extended period of time. Market professionals that are involved in option contracts having the stock support in range bound are likely to lose their investment. The moment of buying a contract should be analyzed carefully, and here is the time when transmission coefficient comes in play.

I will consider here only the case of call option contracts leaving to the interested reader the pleasure of finding alternative investment strategies by applying the transmission coefficient to other type of financial instruments. Still, it should notice that the mathematical reasoning is the same also for put contracts just the stock price is expected to tunnel down.

As is to be expected the basis of the trading strategy is the relation of transmission coefficient (2). A brief look at the equation (2) would tell that the terms trader will deal with are:

- Risk-free interested rate – is a known parameter that usually is taken to be the U.S Treasury rates;
- Distance between resistance and support levels or, in other words the width of the range bound – is also a known parameter and can be easily computed. It can be encountered also in some other trading systems that are based on resistance and support.
- Volatility – is the well-known imply volatility that is computed in a daily basis by all professionals activating in the finance industry. Although is the most important ingredient in the trading system I expose in this paper I will not enter in much details concerning the volatility, assuming that is a concept already entered in the trading routine. A lot of academic and trading literature is devoted to volatility in all its forms. The interested reader not yet familiar with the term the internet is full of resources.

All the components in the relation (3) are known and computing the transmission coefficient is a trivial task. Transmission coefficient should be computed every time imply volatility changes its value.

As empirical tests revel a transmission coefficient T over 95 % is required for the market to be in piece tunneling position. In other words a probability higher than 95 % should be expected before a buy or a sell order is launched in the market.

Some examples of transmission coefficient computing are shown in table 1 below:

Table 1.

For the LNKD, GOOG, HU, NFLX stocks are computed T, the fall in volatility and also the distance of price penetration out of range bound. The price penetration distance has the relation:

$$d = \sqrt{\frac{\sigma}{r}} - K \qquad (4)$$

and is explained in details in [1]. I will only notice here that it represent the minimum of magnitude that the stock price surpass the resistance/support with. It could be chosen to be the exit point for the strategy.

Notice from the table 1 the values of T for the price tunneling to happen and the dramatic short time fall in the stocks volatility prior to price penetration out of the bounded region.

Price tunneling out of the region bounded by levels of support and resistance take the course to this paper to another important notion of the strategy that is choosing the option exercise price as close as possible to support/resistance. This choice for the strike price assures the highest magnitude of price movement and so the optimal level for profits the strategy could promise.

To conclude the strategy discussion the steps in implementing the trading system are recap in the following sentences:

1. Stock market should be in range bound and the distance between the levels of support and resistance should be computed;
2. Risk-free interest rate must be known along with the computing of imply volatility for the financial instrument found in range bound;
3. The transmission coefficient must be computed for every changes in the imply volatility so that a level of T that surpass 95 percent is the expected trigger signal for getting involved in an option contract;
4. Having the T at the appropriate level the market practitioners have to choose the exercise price for the option contract as close as possible to the resistance level, in case of a call and to support level in case of a put.
5. The exit it is leaved in the care of the trader and its behavior vis-à-vis to risk taking. For institutional traders the exit is calibrated according to the risk management. It could be chose the penetration distance d as the exit plan.

The last notice of this section is that the transmission coefficient as it can be easily conclude might be applied in trading all the financial instruments, not only option contracts.

## 5. Conclusions

Recent findings from the paper "Time independent pricing of options in range bound markets" are used in attempt to create wining strategies for option contracts having the underlying stocks in range bounds. Discussion is centered on some particular price movements characterized by an explosive penetration of support or resistance. These brutal price movements with high magnitude are well known by market practitioners as there is widely encountered market phenomenon. The phenomenon is called price tunneling effect.

The probability of such stock price movements in the market is reveled in transmission coefficient equation. Transmission coefficient is design to help market professionals in their investment and risk management decisions by correct timing the moment of entering the market.

The steps of implementing a wining option trading strategy are discussed and some particular examples were given. For some sample stocks transmission coefficient, the penetration distance and the dramatic fall in volatility are computed.

Evidences of price tunneling effect happen in the market are shown in two figures for the LNKD and GOOG stocks. The two figures revealed also the volatility fall prior the price penetration out of the range bound.

It should be noticed that the trading strategy based on transmission coefficient is also functional for a large number of financial instruments, not only for options.

| | date | r | σ | Price at resistance | Price at support | K | d | T | Fall in volatility |
|---|---|---|---|---|---|---|---|---|---|
| **LNKD - LINKEDIN CORP** | 07-08.02.2013 | 0.03 | 0.47 | 127.2 | 123.3 | 3.9 | 0.058114 | 0.998675 | 0.63 to 0.39 |
| **GOOG - GOOGLE INC A** | 22-23.01.2013 | 0.03 | 0.15 | 704.7 | 702.6 | 2.1 | 0.136068 | 0.95 | 0.40 to 0.15 |
| **HUM - HUMANA INC** | 28.03-02.04.2013 | 0.03 | 0.31 | 70.08 | 66.95 | 3.13 | 0.08455 | 0.9948 | 0.43 to 0.25 |
| **NFLX - NETFLIX INC** | 23-24.01.2013 | 0.03 | 0.55 | 101.17 | 97.81 | 3.36 | 0.921744 | 0.933 | 0.95 to 0.55 |

Table 1. Transmission coefficient and fall in volatility prior price tunneling for some stocks.

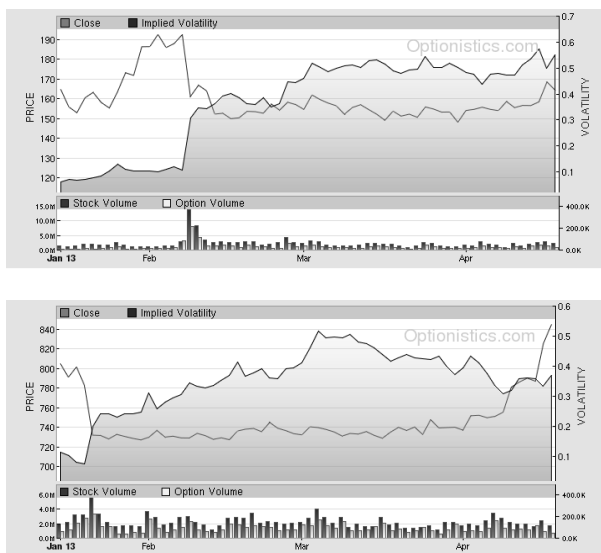

Figure 1. The fall in the stock volatility preceding the tunneling price effect for LNKD (up) and GOOG (down) stocks.

8